\documentclass{article}
\font\openface=msbm10 at10pt
\def\spose#1{\hbox to 0pt{#1\hss}}
\def\lta{\mathrel{\spose{\lower 3pt\hbox{$\mathchar"218$}}
     \raise 2.0pt\hbox{$\mathchar"13C$}}}
\def\gta{\mathrel{\spose{\lower 3pt\hbox{$\mathchar"218$}}
     \raise 2.0pt\hbox{$\mathchar"13E$}}}
\newcommand{\be}{\begin{equation}}
\newcommand{\en}{\end{equation}}
\newcommand{\bea}{\begin{eqnarray}}
\newcommand{\ena}{\end{eqnarray}}

\newcommand{\Tr}{\mbox{Tr}}
\newcommand{\mat}{\mbox{Mat}}
\newcommand{\lap}{{\cal L}^2}

\begin{document}
\title{Fuzzy Orbifolds}
\author{X. Martin}
\maketitle

\begin{abstract} 
A family of fuzzy orbifolds are generated by looking at sub--algebras
of the fuzzy sphere. One of them is actually commutative and can be
mapped exactly onto a lattice. The others are fuzzy approximations of
$S^2/Z_N$ where $Z_N$ is the cyclic group of rotations of angle
$2\pi/N$ and provides the first example of the ``fuzzification'' of a
space with singularities (at the poles). This construction can easily
be generalised to other fuzzy spaces.
\end{abstract}

\section{Introduction}
The idea behind a fuzzy space is to approximate the algebra of
functions of a (commutative) space by a sequence of finite
dimensional algebra, i.e. an algebra of matrices, of increasing
dimension. 

``Fuzzifying'' a space can be done in at least two ways.  The first
one is to quantise the algebra of functions of a commutative space as
a phase space. The archetypal example of this method is the fuzzy
sphere \cite{Madore}, but complex projective spaces $\mbox{\openface
CP}^n$ and other coadjoint orbits can also be treated this way. In
this case, the algebra of matrices can be embedded into the algebra of
functions of the commutative space with a deformed product called the
star product \cite{BBDJX}, the symmetries of the commutative space are
exactly preserved on the matrix approximations, and derivations can be
identified with infinitesimal transformations of the symmetry
group. From there, any algebraic expression in the algebra of function
can easily be approximated on the matrix algebras.

We are interested in field theories, and this kind of fuzzy space can
then be used for non--perturbative studies \cite{xavier}. The space
can be described by a triple containing a sequence of algebras which
reduces essentially to a sequence of dimensions of matrix spaces, a
differential operator necessary to describe the kinetic part of the
action (Laplacian for a scalar field, Dirac operator for a spinor
field...) and a scalar product to produce a scalar action. With these
ingredients, it is possible to describe field theories on the fuzzy
space.

This is a very elegant method, but unfortunately it can only be
applied to a limited number of spaces. Note however, that Cartesian
products and ``fuzzification'' commute allowing one to ``fuzzify'' any
Cartesian product of such spaces easily.

The second ``fuzzification'' method only requires that a functional
integral on the commutative space be approximated by an integral on
the matrices. This is a much weaker requirement which is sufficient
for the purpose of studying field theory, and allows one to ``fuzzify'' a
much larger class of spaces \cite{bd}. The trick is generally to
select a linear sub--space with the right finite dimensional
representation of the symmetry group, and then perform a kind of
compactification on the additional degrees of freedom by putting an
addition weight in the functional integral. Again, the necessary
derivation operators can be associated with infinitesimal
transformations of the symmetry group.  In this case, the triple can
also be defined, but does not describe all the structure since the
functional integration measure must also be added to it.

Although this latter method allows for a much larger class of spaces,
even including spaces with edges \cite{fdisk}, it would
appear to be limited to smooth, non--singular, spaces.

In this paper, we observe that there are sub--algebras of the fuzzy
sphere stable under the Laplacian operator, which can therefore be
used to ``fuzzify'' scalar field actions on subspaces of the algebra
of functions. In this case, this sub--algebra can be identified with the
algebra of functions of the orbifold $S^2/Z_n$ with $Z_n$ the cyclic
group with $N$ elements. Similar results can certainly be generalised
to more complicated fuzzy spaces such as the complex projective planes
$\mbox{\openface CP}^n$ without difficulty.

In the next Section, we introduce the fuzzy sphere before introducing
in Section three its sub--algebras which give rise to the fuzzy
orbifolds. These sub--algebras are actually split between a
commutative sub--algebra which, together with its Laplacian, can be
mapped onto a standard lattice, and non--commutative
sub--algebras. Finally, we conclude in Section four.

\section{The fuzzy sphere}
The simplest example of a fuzzy space is the fuzzy sphere
\cite{Madore}. As explained in the Introduction, for the purpose of
studying a scalar field theory, the only ingredient required to fix
the geometry is a Laplacian operator and a scalar product on each
matrix algebra. Since derivations on the commutative sphere can be
viewed as infinitesimal $\mathrm{SU}(2)$ transformations, the
Laplacian on a $(2s+1)\times (2s+1)$ matrix algebra, denoted
$\mat_{2s+1}$, is proposed as \be 
\lap \phi=[L_i,[L_i,\phi]], \label{flap} \en 
where $L_i$ are the angular momentum operators in the $2s+1$
dimensional irreducible representation of $\mathrm{SU}(2)$. The
canonical matrix scalar product \be
<\phi|\psi>=\frac{4\pi}{2s+1} \Tr(\phi^\dag \psi). \label{scpr} \en
is chosen with a multiplicative coefficient such that the unit matrix
has the same norm as the unit function on the sphere of radius one.

The spectrum and eigenmatrices of the proposed Laplacian can be
recognised from the adjoint action of angular momentum as \bea 
\lap \hat{Y}_{lm} & = & l(l+1)\hat{Y}_{lm} ,\ 0\leq l\leq 2s
\label{fsh} \\ {[} L_3, \hat{Y}_{lm} ] & = & m\hat{Y}_{lm} ,\ 0\leq
|m|\leq l, \label{L3Ylm} \ena
where the matrices $\hat{Y}_{lm}$ are the polarisation tensors
 whose normalisation is defined according to the chosen scalar product \be
\frac{4\pi}{2s+1} \Tr(\hat{Y}_{lm}^\dag \hat{Y}_{lm})=1.\en
This is precisely the spectrum of the Laplacian on the commutative
sphere truncated at angular momentum $2s$, thus vindicating this
choice.

A clean way of recognising the approximation of a sphere in these
matrix algebras is to introduce a mapping which associates a function
on the sphere with each matrix of the algebra $\mat_{2s+1}$ and pulls
back most of the structure on the algebra of functions of the sphere
onto the matrix algebra. There are various ways to define such a
mapping, such as using coherent states \cite{pres} or the Brezin
symbol map \cite{denjoemap}. For instance, the latter is given by \bea 
\mathcal{M}_s: \mat_{2s+1} & \rightarrow & \mathcal{C}^\infty
(S^2) \\ M=\sum_{l=0}^{2s}\sum_{m=-l}^l c_{lm}\hat{Y}_{lm} & \mapsto &
f({\bf n})=\sum_{l=0}^{2s}\sum_{m=-l}^l c_{lm} Y_{lm}({\bf n}),
\label{map} \ena 
where the functions $Y_{lm}({\bf n})$ are the usual spherical harmonics
on the sphere, i.e. the eigenvectors of the Laplacian operator on the
sphere. By definition, this mapping $\mathcal{M}_s$ is linear and maps
the Laplacian $\lap$ on $\mat_{2s+1}$ onto the Laplacian on the
sphere. In fact, the three
derivatives on the sphere $\nabla_l=i\varepsilon _{jkl} x_j
\partial_k$ are pulled back to simple derivations on the matrix
algebra given by \be
\mathcal{L}_i\phi=[L_i,\phi ] .\label{fsphder} \en
By construction, the action of the group $\mathrm{SU}(2)$ is preserved on both
sides. Furthermore, since the eigenvectors of the Laplacian on the
matrix space and on the sphere form orthonormal bases on their
respective spaces, this mapping is an injective isometry. Its image,
on which the mapping is one to one, $\mathcal{M}_s(\mat_{2s+1})$ is
given by all the functions with angular momentum only up to $2s$ and
form a sequence of increasing (for the inclusion) sets which become
dense in $\mathcal{C}^\infty (S^2)$ in the limit of infinite
matrices. The matrix product is mapped to a (non--commutative) product
of functions on the sphere called a $*$--product \be
\mathcal{M}_s (\phi\psi)({\bf n})=(\mathcal{M}_s(\phi)*_s
\mathcal{M}_s(\psi)) ({\bf n}),\en
which is evidently distinct from the usual (commutative) product of
functions. It is possible to verify that in the limit of infinite
matrices $s\rightarrow \infty$, the star product tends to the usual
product. More precisely, for $(f_s,g_s)\in (\mathcal{M}_s(\mat
_{2s+1}))^2$ two functions with angular momentum truncated at $2s$,
and $t\geq s$, \be 
(f_s*_t g_s)({\bf n})=f_s({\bf n})g_s({\bf n})+\mathcal{O}(
\frac{1}{t}).\en

Note in passing that complex conjugation of a function on the sphere
pulls back to hermitian conjugation on the matrix algebra.
Consequently, as proposed in the introduction, real functions pull
back to hermitian matrices. Similarly, integration on the sphere
which is similar to scalar product with the unit function pulls back
to the trace on the matrix algebra.

Thus, in the limit when $s$ goes to infinity, the mapping
$\mathcal{M}_s$ becomes an isomorphism of algebras which preserves
rotational invariance, the Laplacian and the scalar product
(\ref{scpr}). This proves that the fuzzy spaces, as defined by the
triple $(\mat_{2s+1},\lap,<\cdot,\cdot>)$ go over to the sphere in
the limit of infinitely large matrices.

Another mapping with similar properties which is often introduced is
the one obtained by looking at the diagonal elements of a matrix in a
coherent states representation. Compared to $\mathcal{M}_s$, this
mapping trades the isometry property for the conservation of the
notion of state, in the sense that it maps a state of $\mat_{2s+1}$
into a state of $\mathcal{C}^\infty (S^2)$. More generally it
conserves the notion of positivity in the sense that a positive matrix
is sent to a positive function. In this case, the corresponding star
product can also be expressed in a simple exact form \cite{pres}.

This introduction to the fuzzy sphere described spheres of radius
one. Getting spheres of different radius $R$, is just a matter of
scaling the scalar product (\ref{scpr}) and Laplacian (\ref{flap})
appropriately: \bea 
\lap & \rightarrow & \frac{1}{R^2} \lap ,\label{scaleL} \\ 
\frac{4\pi}{2s+1} \Tr(\phi^\dag \psi) & \rightarrow & \frac{4\pi
  R^2}{2s+1} \Tr(\phi^\dag \psi) .\label{scaletr} \ena 
Since this is such a simple generalisation, only spheres of radius one
will be considered in the following.

With the fuzzy sphere cleanly defined, its sub--algebras can now be
investigated. 

\section{sub--algebras of the fuzzy sphere}
As mentioned and illustrated in the introduction in the case of the
sphere, a triple of a sequence of algebras of diverging dimension, a
Laplacian and a scalar product is sufficient to define a fuzzy space
on which a scalar field field theory can be studied.

It is clear that if a sequence of matrix sub--algebras stable under
the action of the Laplacian can be found, a similar triple will
immediately be induced on them. Stability under hermitian conjugation
is not necessary but will also be retained in our examples. In a way,
the construction here can be viewed as a method of defining a ``fuzzy
sub-space''.

In the case of the fuzzy sphere, such a sub--algebra must be generated
by a family of eigenvectors of the Laplacian $\hat{Y}_{lm}$ closed
under multiplication and hermitian conjugation, or equivalently
under $m\rightarrow -m$.

\subsection{The commutative sub--algebra}
The simplest such example one can imagine is given by the sub--algebra
of diagonal matrices which will be called ${\cal A}_0^s$. These are
generated exactly by the polarisation operators $\hat{Y}_{l0}$. To see
this, note first that these two spaces have the same dimension $2s+1$,
and that the $\hat{Y}_{l0}$ are themselves diagonal since, using
(\ref{L3Ylm}),  for $i\not=j$, \be
(\hat{Y}_{l0})_{ij}=\frac{1}{i-j}({\cal L}_3 \hat{Y}_{l0})_{ij}=0.
\label{Ytrick} \en
Thus, in the limit where $s$ tends to infinity, the mapping ${\cal
M}_s$ defined in Eq. (\ref{map}) sends this algebra ${\cal A}_0^s$
to the algebra of functions ${\cal A}_0^\infty$ on the sphere
invariant under rotations around the third axis \be
{\cal A}_0^\infty=\{ \phi(\theta,\phi)|\phi(\theta,\phi)=\phi(\theta)
\} ,\label{fuzsub0} \en
and using the usual Laplacian on the sphere \be
-\Delta \phi= \frac{d}{\sin(\theta)d\theta}(\sin(\theta)\frac{d\phi}{d
\theta}) \label{Laplsph} \en
where $(\theta,\phi)$ are the spherical coordinates. Equivalently, by 
changing variables from $\theta$ to $z=\cos(\theta)$, this space can
be viewed as the space of ${\cal C}^2$ functions on the segment
\be V_0=[ -1,1] ,\label{segment} \en
with the Laplacian \be
\Delta f =\frac{d}{dz}((1-z^2)\frac{d\, f}{dz}),\label{lapd} \en
and no constraints at the boundaries $z=\pm 1$.

In conclusion, the sequence of matrix sub--algebras ${\cal A}_0^s$ is
a ``fuzzy subspace''of the fuzzy sphere which approximates a segment
with the Laplacian (\ref{lapd}).

Since the algebra ${\cal A}_0^s$ is actually commutative, it should be
possible to map it to an algebra of functions of a lattice. To see
that, the degrees of freedom of the algebra ${\cal A}_0^s$ must be
identified with some lattice points, and the Laplacian with a finite
difference Laplacian. The Laplacian on the discrete algebra \be
{\cal A}_0^s=\{ \mbox{Diag}(\phi_i)_{-l\leq i\leq l}|\phi_i \in
\mathbf{C} \} ,\en
is given by the diagonal matrix \be
-({\cal L}^2 \phi)_{ii}=(l(l+1)-i^2)(\phi_{i+1}+\phi_{i+1}-2\phi_i)
-i(\phi_{i+1}-\phi_{i+1}).\label{fuzlapd} \en
It is clear that this Laplacian (\ref{fuzlapd}) is a finite difference
approximation of the commutative Laplacian (\ref{lapd}) on the lattice
where $z=i/\sqrt{l(l+1)}$, that is \be
\phi_i=\phi(\arccos(i/\sqrt{l(l+1)}), \ -l\leq i \leq l.\en

Note that the sub--algebras ${\cal A}^s_0$ have no continuous symmetry
left, and that the derivations defined on the fuzzy sphere by
(\ref{fsphder}) do not close in ${\cal A}_0^s$ except for ${\cal L}_3$
which is trivially zero. This is as it should be for a lattice.

Further sub--algebras of ${\cal A}^s_0$ can easily be constructed by
considering it as a lattice. However, they hold little interest from
the point of view of fuzzy spaces.

\subsection{Other orbifolds}
More generally, the product of two polarisation operators can be
expanded on the basis of polarisation operators itself in a
complicated way which involves Clebsh--Gordan coefficients. However,
it should be noted that the Leibnitz rule implies that the axial
quantum number is just added under multiplication, that is \be {\cal
L}_3 (\hat{Y}_{lm}\hat{Y}_{kn})=(m+n)\hat{Y}_{lm}\hat{Y}_{kn} .\en

Thus a generalisation of the sub--algebra ${\cal A}_0^s$ is given by the
sub--algebras ${\cal A}_k^s$, $k$ a fixed positive integer, generated by 
the polarisation operators with axial quantum number a multiple of $k$
\be {\cal A}_k^s=\mbox{Span}(\hat{Y}_{l\, km}, \ 0\leq l \leq s,\
0\leq k|m| \leq l).\label{fuzsubk} \en 
The notation ${\cal A}_k^s$ adopted here is consistent with that of  
the space ${\cal A}^s_0$ introduced in the previous subsection which
does indeed correspond to the case when $k=0$. More generally, if the
requirement that the space be invariant under conjugation is dropped,
spaces such as \be
{\cal A}_k^{\pm s}=\mbox{Span}(\hat{Y}_{l\, km}, \ 0\leq l \leq s,\
0\leq \pm km \leq l), \en
are also acceptable.

Using the same trick as in Eq. (\ref{Ytrick}), it is possible to check
that $\hat{Y}_{l\, km}$ has only non--zero entries on the
off--diagonal $(i,i-m)$. Thus, the sub--algebra ${\cal A}_k^s$ can be
recognised as the space of band diagonal $(2s+1)\times (2s+1)$
matrices which are non--zero on the diagonal and then alternate $k-1$
diagonals of zeros with a non--zero diagonal, or written
algebraically, \be
{\cal A}_k^s=\{ M\in \mbox{Mat}_{2s+1}| M_{ij}=0,\ i\not\equiv
j\mbox{mod} k \}.\en
Again, these sequences of sub--algebras can be seen as fuzzy
approximations of the algebra of functions on the sphere generated by
the family of spherical harmonics $Y_{l\; km}$. These in turn can be
identified with the space of functions invariant under rotation around
the third axis of angle $2\pi/k$, or conversely as the space of
functions on the slice of the sphere defined by \be
V_k =\{ (\sin(\theta)\cos(\phi),\sin(\theta)\sin(\phi),\cos(
\theta))| \ 0\leq \phi<2\pi/k \},\en
with periodic boundary conditions between $\phi=0$ and $\phi=2\pi/k$.
This space of functions is equivalent to the orbifold $S^2/Z_k$, with
$Z_k$ the cyclic group with $k$ elements.

Note that in this case as in the commutative case, the space ${\cal
A}_k^s$ is not closed under derivations ${\cal L}_i$ as defined in
(\ref{fsphder}) unless $k=1$ which corresponds to the fuzzy sphere. On
the other hand, this space retains a continuous $U(1)$ symmetry group
corresponding to axial rotation around the thirs axis.

Of course, the algebra ${\cal A}_k^s$ can be rewritten as a direct sum
of matrix algebras itself \be
{\cal A}_k^s=\oplus_{i=0}^{k-1} E_i\otimes E_i ,\ E_i=\oplus_{0\leq
  k|m|+i \leq l} |s,km+i>,\label{blockm}\en  
where $|s,m>$ with $0\leq |m| \leq 2s+1$ is the canonical basis of the
linear space on which $(2s+1)\times (2s+1)$ matrices act.

There does not appear to be other such examples on the fuzzy
sphere. For instance, ${\cal RP}^2$, whose algebra of function is
generated by a family of spherical harmonics, the $Y_{2l\; m}$, can
not be fuzzified in this way because the corresponding family of
polarisation operators, $\hat{Y}_{2l\; m}$ is not stable under
multiplication.

\section{Conclusion}
A family of ``fuzzy sub--spaces'' of the fuzzy sphere were proposed
which are sub--algebras closed under the Laplacian so that a triple is
automatically induced from the triple of their parent space. There is
an obvious generalisation of this construction to other fuzzy spaces
such as the complex projective planes. Although the fuzzy spheres of
other dimensions and tori introduced in \cite{bd} where constructed in
a different way than the fuzzy sphere, a similar construction can also
be performed there.

The main difference with previously proposed fuzzy spaces is that the
limiting commutative space is singular (at the poles) and that the
algebra is a direct sum of matrix algebras instead of being a single
matrix algebra. This suggests a possible relation between these two
properties.

\end{document}